 \documentclass[prl,aps,twocolumn,showpacs]{revtex4}
\usepackage[dvips]{graphicx}
\usepackage{dcolumn}
\newcommand{\be}{\begin{equation}}
\newcommand{\ee}{\end{equation}}
\newcommand{\bea}{\begin{eqnarray}}
\newcommand{\eea}{\end{eqnarray}}

\begin{document}
\topmargin=-18mm

\title{ Long-Range Electron Transfer  and Electronic Transport Through Macromolecules}

\author{ Natalya Zimbovskaya$^{\dagger\ddagger} $  and Godfrey Gumbs$^\ddagger $ }
\vspace{2mm}

\address{$^{\dagger} $Department of Physics 
City College of CUNY,
Convent Avenue \& 138th Street, New York, NY 10031, 
\\$^\ddagger$Department of Physics and Astronomy
Hunter College of CUNY,
695 Park Avenue, New York, NY 10021}

\begin{abstract} 
 A theory of electrical transport through molecular wires is used to
estimate the electronic factor in the intramolecular electron transfer
(ET) in porphyrin-nitrobenzene supermolecules, and to analyze its
structure.  The chosen molecules have complex donor and acceptor
configurations, and relatively simple structure of the bridge, which
enables us to concentrate our studies on donor/acceptor coupling to the
bridge.  We present analytical and numerical results concerning the effect
of donor/acceptor coupling to the bridge on the ET process in molecules
with complex donor/acceptor subsystems.
 \end{abstract}

\maketitle


Long-distance electron transfer (ET) plays a central role in many biological processes \cite{1}. It has been established that molecular ET is essentially a combination of nuclear environment fluctuations and electron tunneling \cite{2}. The expression for the ET rate including both electronic and nuclear factors was first proposed by Marcus \cite{3,4,5}. The existing theories of long-range ET are mainly concentrated on the analysis of the electron tunneling through the bridge (Refs \cite{6,7,8,9,10,11,12}). Hitherto, less attention has been paid to systematic studies of donor/acceptor coupling to the bridge and its contribution to the electronic transmission coefficient. Usually, theoretical models of ET present both donor and acceptor as single sites with one state per site. In realistic biological systems, however, donor and acceptor subsystems are complex and include numerous sites which provide effective  coupling to the bridge.

In this letter, we study the effect of the donor/acceptor subsystems on the characteristics of the long-range ET in macromolecules. We use an apparent close similarity between this process and an electron transport through molecular wire placed between metallic contacts. This enables us to apply the methods developed in Refs. \cite{13,14,15,16} to calculate the electron transmission function $ T(E) $ which we can further analyze to separate out contributions from different donor/acceptor sites participating in the ET process. We also calculate current-voltage characteristics of the intramolecular electron transport.

In further studies, we simulate our donor and acceptor subsystems as sets of semi-infinite tight-binding homogeneous chains. Each chain is attached to a site of donor/acceptor  which can be effectively coupled to the bridge. To simplify the following calculations we assume that the
chains  do not interact with each other. This model for the donor/acceptor subsystem is a generalization of that adopted in the theory of electron transport through molecular wires \cite{13,14,15,16} where each metallic electrode was treated as a single chain.  However, in the present analysis
we need this generalization to take into account the complexity of the donor/acceptor subsystems.

To proceed we consider each element of the donor-bridge-acceptor system as an array of potential wells.  Each well represents one atomic orbital of the corresponding part of the molecule (donor, acceptor or bridge).  These wells are sites of the donor-bridge-acceptor system.  Within this
approach, any atom will be represented by a set of sites corresponding to its states, and we do not distinguish between sites corresponding to the same atom, and those  corresponding to different atoms of our molecular system \cite{8,17}.

In the following calculations we start from a tight-binding Hamiltonian  for the bridge:
  \begin{equation} 
H = H_0 + H_1
   \label{e1}
       \end{equation}
 whose matrix elements between states $|i>$ and $|j>$ corresponding to the $i$-th and $j$-th sites are given by:
 \begin{equation} 
\left(H_0 \right)_{ij}= \alpha_i\delta_{ij} \  ; \qquad
\left(H_1 \right)_{ij}=  {\cal V}_{ij}\ ,
     \label{e2}
           \end{equation}
 where $ \delta_{ij} $ is Kronecker symbol, ${\cal V}_{ij}$= 0 when $i=j$ and only states associated with  valence electrons are considered.  The diagonal matrix elements $\alpha_i$ are 
ionization energies of electrons at sites $i$, while the off-diagonal matrix elements ${\cal V}_{ij} = {\cal V}_{ji}$ are the hopping strengths between the $i$th and $j$th sites. Both direct and exchange energy contribution are included in $ {\cal V}_{ij}.$  As well as it was carried out in \cite{15,16} we also take into account self-energy corrections arising due to the coupling of the donor $\left(\Sigma_{\cal D}\right)$ and acceptor $\left(\Sigma_ {\cal A}\right)$ to the bridge. As a result, we arrive at the effective Hamiltonian 
for the bridge, i.e.,
  \begin{equation} 
H_{eff}= H + H_{\cal D} + H_{\cal A}\ .
\label{e3}
       \end{equation}
 Generalizing the results of Ref. \cite{18} for the case when we have several donor and acceptor sites coupling to the bridge, we obtain
  \begin{equation}
(H_{\cal D})_{ij} = \Big(\Sigma_D  \Big)_{i} \delta_{ij}
 \equiv \sum_k\frac{{\cal D}_{ki}^{\, 2}}{E-\epsilon_k-\sigma_k} \delta_{ij} \ ,
      \label{e4}
         \end{equation}
  \begin{equation}
(H_{\cal A})_{ij} =\left(\Sigma_A  \right)_i \delta_{ij}
\equiv \sum_l\frac{{\cal A}_{il}^{\, 2}}{E-\epsilon_l-\sigma_l} \delta_{ij}\ .
  \label{e5}
        \end{equation}
  Here, ${\cal D} _{ki}$ and ${\cal A}_{il}$ are, respectively, coupling strengths between the $k$-th donor site or the $l$-th acceptor site and the $i$-th site of the bridge, and $\sigma_{k,l} = \frac{1}{2} \left\{\theta_{k,l} - i \sqrt{4 \gamma_{k,l}^2 - \theta_{k,l}^2}\right\}$ are the self-energy corrections of the semiinfinite chains attached to the corresponding sites \cite{13}.  The parameters $\theta _{k,l} = E - \epsilon_{k,l}$, $\epsilon_{k,l}$ and $\gamma_{k,l}$ are the ionization
energies of electrons at the corresponding donor/acceptor sites, and the nearest-neighbor hopping integrals for the chains.  Summation in Eqs. (\ref{e4}) and (\ref{e5}) is carried out over all donor/acceptor sites coupled to the bridge. 
Due to the presence of the self-energy corrections $\Sigma_{\cal D, \cal A}$, the eigenvalues of the effective Hamiltonian (\ref{e3}) include imaginary parts ${\Gamma}_i$ which represent broadening of the bridge energy levels $E_i$ originating from the coupling of the bridge to the donor and acceptor systems. The energy levels are broadened further if
include scattering processes in the bridge.  Here, however, for
simplicity, we separate the effect of electronic dephasing, assuming that coherent electronic tunneling through a bridge to be the predominant mechanism of ET.

Treating the long-range ET processes, we can neglect the direct coupling of donor sites to acceptor sites.  Then the probability amplitude for the transition between the $k$-th donor site and the $l$-th acceptor site is given by \cite{19}
      \begin{equation}
T_{kl}= \sum_{i,j}{\cal D}_{ki} {\cal G}_{ij}{\cal A}_{jl}\ .
    \label{e6}
                \end{equation}
 Here, the sum is carried out over the bridge sites, $G_{ij}$ is the matrix element of the Green's function corresponding to $H_{eff}$, i.e.,
  \begin{equation}
{\cal G}_{ij}=\left<i\left|\left( E-H_{eff} \right)^{-1} \right|  j\right>\ .
       \label{e7}
             \end{equation}
 An electric tunneling current $I$ flowing from donor to acceptor through the bridge in the presence of a small applied voltage $V$ has the form \cite{20}
  \begin{equation} 
I= \frac{2 \pi e}{\hbar} \sum_{k,l} f \left( E_k \right) \lbrack 1- f \left( E_l +eV \right)\rbrack \mid T_{kl}^2 \mid \delta \left(E_k - E_l \right)
 \label{e8}
          \end{equation}
 where $f(E)$ is the Fermi function; $e$ is the charge of the electron.  Assuming that the applied potential varies linearly in the molecule, we can use the approximation $ V = V_0 (L_b /L_{mol} ),$ where $ L_b $ and $ L_{mol} $ are the lengths of the bridge and the whole molecule, respectively, and $ V_0 $ is a voltage applied across the whole molecule. Starting from the expression (8) and following a usual way we arrive at the
standard formula  \cite{21}:
   \begin{equation}
I = \frac{e}{\pi \hbar}\int_{- \infty}^{ \infty}dE\ T(E) \lbrack f \left( E - {\mu}_1
\right)- f \left( E - {\mu}_2 \right) \rbrack\ ,
 \label{e9}
        \end{equation}
 where the chemical potentials ${\mu}_1$ and ${\mu}_2$ are determined by the equilibrium Fermi energy of the bridge $E_F$ and the effective voltage $V$ across the bridge \cite{16}, i.e.,
    \[
{\mu}_1 = E_F + \left( 1 - \eta \right) eV;
\qquad
  {\mu}_2 = E_F - \eta e V. \] 
 Here the parameter $\eta$ characterizes how the voltage $V $ is divided between the two ends of the bridge. The equilibrium Fermi energy for the bridge can be determined from the equation (See Ref. \cite{16}):
  \begin{equation} 
N = \frac{2}{\pi}\sum_i \left( \frac{\pi}{2}+ \arctan \frac{E_f - E_i}{\Gamma_i}  \right)\ ,
   \label{e10}
      \end{equation}
 where  $N$ is the number of electrons on the bridge part of the molecule.

The electron transmission function included in the expression (9) is given by the formula   
  \begin{equation} 
T(E) = 2 \sum_{i,j} \Delta_i |G_{ij}|^2 \Delta_j \ ,
   \label{e11}
      \end{equation}
 where the summation is carried out over the states of the bridge, and the quantities   $\Delta_{i,j}$ are defined as
   \[
\Delta_i = \mbox{Im} \left(\Sigma_{\cal D}\right)_{ii}, \qquad
\Delta_j = \mbox{Im} \left(\Sigma_{\cal A}\right)_{ii}\ .
   \]
 This formula agrees with the result obtained earlier in the study of electron transport through molecular wires \cite{15}.  
 
 \begin{figure}[t] 
\begin{center}
\includegraphics[width=6.0cm,height=6cm]{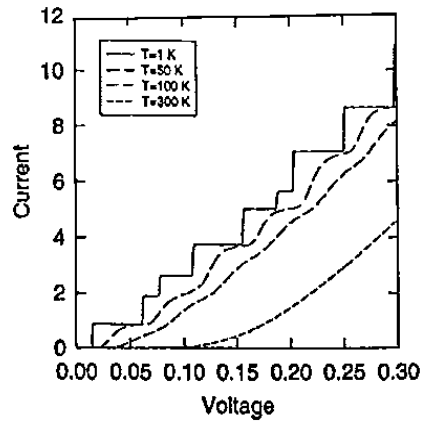}
\caption{
The calculated current (nA) -- voltage (V) characteristics for different temperatures.}   
\label{rateI}
\end{center}
\end{figure}

We see that contributions to the transmission function $T(E)$ from different donor and acceptor sites are determined by the factors $\Delta_{i,j}$.  It follows from Eqs. (\ref{e4}) and (\ref{e5}) that these factors strongly depend on energy-dependent denominators.  When the tunneling energy $E$ takes on a value close to $\epsilon_{k,l}$, the corresponding term in Eqs.\ (\ref{e4}) and (\ref{e5}) can surpass all remaining terms.  Thus, for different values of $E$, different donor and
acceptor sites can predominate in the ET process.  Expressions (\ref{e4}) and (\ref{e5}) enable us to determine these dominating sites for any interval of tunneling energy $E$.
 In Fig.1 we present results for the calculated I--V characteristics for the porphyrine-nitrobenzene macromolecules for different temperatures (T = 1K, 50K, 100K, 300K). In calculations we estimate $ E_F $ using the expression (10) and we choose $ \eta = 0.3. $ Matrix elements of the bridge Hamiltonian (1) as well as the ionization energies for the donor/acceptor
sites are obtained from the extended Huckel method.
The  parameters $ \gamma_{k,l}$ representing hopping integrals for donor/acceptor chains are estimated as averages of matrix elements which correspond to the coupling of these sites to their nearest neighbors within the donor/acceptor part of the molecule. 

\begin{figure}[t] 
\begin{center}
\includegraphics[width=6.0cm,height=6.0cm]{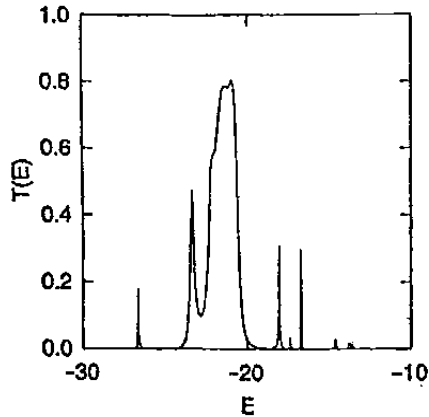}
\caption{
The calculated electron transmission function through the bridge $ T(E)$ within a certain range of energies $ E(eV)$.} 
\label{rateI}
\end{center}
\end{figure}
Comparison of our results with those obtained for I--V curves
characterizing the electric transport through molecular wires \cite{14} shows their resemblance at room temperatures. At low temperatures we notice that our I--V curves exhibit a step-like behavior and the plateaus get narrower as the temperature raises. Associating the low-temperature I--V curves with the result for the electron transmission function $T (E)
$ shown in part in Fig. 2 we see that the steps reflect the structure of $T(E)$. The latter is a set of peaks separated by the intervals where the electron transmission function takes on values close to zero. These peaks arise due to the contributions from different donor-acceptor sites as well
as from the Greens function corresponding to the effective Hamiltonian for the bridge. At a given voltage $V$ the difference of the Fermi functions in expression (9) differs from zero only in the interior of a certain energy range including $E_F$. Therefore, the tunneling current $I$ is fully determined by the contributions of peaks of the transmission function located there.  When we increase the voltage applied across the bridge the width of the relevant energy range enhances, and extra peaks of the function $ T(E)$ contribute to the current. When a new peak appears this gives rise to a sudden enhancement of $I$, so that the step exhibited in the $I-V$ curves reflect the structure of $T(E) .$  At higher temperatures Fermi distribution functions in the integrand of (9) lose their step-like character and the plateaus are washed out. Another reason for peaks in the electron transmission function (and  the steps in the I--V curves, respectively) to be eroded is the electronic phase-breaking effect which arises in complex molecules due to stochastic fluctuations of ion potential. We conjecture, however, that at low temperatures the electronic dephasing reduces to same extent, so that the structure of $ T(E) $ can be revealed.

In summary, we have presented a theoretical model for the intramolecular ET which resembles those used to study the electric transport through molecular  wires. Our model enables us to analyze in detail the donor/acceptor coupling to the molecular bridge which is important for further  investigations of long-range ET through  macromolecules.

{\bf Acknowledgments.} We thank D. Akins for providing us the data concerning the structure of porphyrin-nitrobenzene molecules, and J. Malinsky for helpful discussions.  GG acknowledges the support in part from grant \# 4137308-02 from the NIH. NAZ also thanks G.M. Zimbovsky for the help with the manuscript.


\begin{references}

\bibitem{1}	A.M. Kuznetsov and I. Ulstrup "Electron Transfer in Physics 
and Biology", Wiley, England, (1999).

\bibitem{2}	H.M. McConnel, J. Chem. Phys. {\bf 35}, 508 (1961).

\bibitem{3}	R.A. Marcus, J. Chem. Phys. {\bf 24}, 979 (1956); 
{\bf 43}, 679 (1965); 
Annu. Rev. Phys. Chem. {\bf 15}, 155 (1964).

\bibitem{4}	N. Sutin, Acc. Chem. Res. {\bf 15}, 275 (1982); Progr. Inorg.
Chem. 
{\bf 30}, 441 (1983).

\bibitem{5}	R.A. Marcus and N. Sutin, Biochem. Biophys. Acta {\bf 811}, 265
(1985)

\bibitem{6} D.N. Beartan, J.N. Betts and J.N. Onuchik, Science {\bf 252},
128 (1991).

\bibitem{7}	J.N. Onuchik, D.N. Beratan, J.R. Winkler and H.B. Gray, 
Science {\bf 258}, 1740 (1992).

\bibitem{8}	C. Goldman, Phys. Rev. A {\bf 43}, 4500 (1991).


\bibitem{9}	S.S. Skorits, J.J. Regan and J.N. Onuchik, J. Phys. Chem. 
98, 3379 (1994).

\bibitem{10} J.J. Regan, A.J. DiBiblo, R. Langen, L.K. Skov, J.R. Winkler,
H.B. Gray, J.N. Onuchik, Chem. Biolog. {\bf 2}, 484 (1995).

\bibitem{11} J.N. Gehlen, I. Daizabeth, A.A. Stuchebrukov and R.A.  
Marcus, Inog. Chem. Acta 243, 271 (1996).


\bibitem{12}	J. Kim and A.A. Stuchebrukhov, J. Phys. Chem. {\bf 104}, 
8608 (2000).


\bibitem{13}	J.L. D'Amato and H.M. Pastawski, Phys. Rev. B {\bf 41},
7411
(1990).

\bibitem{14}	X.-Q. Li and Y. Yan, J. Chem. Phys. {\bf 115}, 4169
(2001); Appl. 
Phys. Lett. {\bf 79}, 2190 (2001).


\bibitem{15}	M.P. Samanta, W. Tian, S. Datta, J.H. Henderson and C.P. 
Kubiak, Phys. Rev. B {\bf 53}, R7626 (1996).

\bibitem{16}	S. Datta, W. Tian, S. Hong, R. Reinfenberger, J.H. Henderson 
and C.P. Kubiak, Phys. Rev. Lett {\bf 79}, 2530 (1997).

\bibitem{17}	J. Malinsky and Y. Magarshak, J. Phys. Chem. 96, 2849 (1992); 
Y. Magarshak, J, Malinsky and A.D. Joran, J. Chem. Phys. {\bf 95}, 418 (1991).
 

\bibitem{18}	V. Mujica, M. Kemp and M.A. Ratner, J. Chem. Phys. 
{\bf 101}, 6849 (1994).
	
\bibitem{19} See, e.g., E.N. Economou, "{\em Green's Functions in Quantum
Physics}", (Springer-Verlag, Berlin, 1979).

\bibitem{20}	J. Bardeen, Phys. Rev. Lett. {\bf 6}, 57 (1961); J. Tersoff and 
D.R. Hamman, Phys. Rev. B {\bf 31}, 805 (1985).

\bibitem{21} S. Datta "{\em Electronic Transport in Mesoscopic Systems}"
(Cambridge University Press, Cambridge, U.K., 1995).

\bibitem{22} Expression (\ref{e10}) is a generalization of the result  
obtained in \cite{18} in the low-temperature limit, and assuming that both
donor and acceptor can be simulated as a single tight-binding chain.


\end{references}
\end{document}